\documentstyle[aas2pp4,epsf]{article}		
\def \sun {\hbox {$\odot$}}

\received{}
\accepted{}

\lefthead{Ba\l uci\'nska-Church et al.}
\righthead{Cessation of Flickering in Cyg X-1}

\begin{document}

\title{The Cessation of Flickering during Dips in Cygnus X-1}
\author{M. BA{\L }UCI\'NSKA-CHURCH\altaffilmark{1,2}, 
T. TAKAHASHI\altaffilmark{2}, Y. UEDA\altaffilmark{2}, 
M. J. CHURCH\altaffilmark{1,2}, \hbox {T. DOTANI\altaffilmark{2}}, 
K. MITSUDA\altaffilmark{2} AND H. INOUE\altaffilmark{2} }

\altaffiltext{1}{School of Physics and Space Research, University of
Birmingham, Edgbaston, Birmingham B15 2TT, UK}
\altaffiltext{2}{Institute of Space and Astronautical Science, Yoshinodai
3-1-1, Sagamihara, Kanagawa 229, JAPAN}


\begin{abstract}
We report the discovery of the cessation of flickering in dips
in the black hole candidate Cygnus X-1, detected for the first time
in the ASCA observation of May 9th., 1995. During this observation,
particularly deep dipping took place resulting in strong changes 
in hardness ratio corresponding to absorption of the power law spectral
component. The deadtime corrected
light curve with high time resolution clearly shows a dramatic
decrease in the extent of flickering in the band 0.7 - 4.0 keV
during dipping, but in the band 4.0 - 10.0 keV, there is relatively
little change. We show that the rms flickering amplitude in the band
0.7 - 4.0 keV is proportional to the X-ray intensity in this band
which changes by a factor of almost three. This is direct evidence that 
the strong Low State flickering is intrinsic to the power law emission;
ie takes place as part of the emission process. The rms amplitude is
proportional to the intensity in the low energy band, except for a
possible deviation from linearity at the lower intensities. If confirmed,
this non-linearity could imply a process such as electron scattering of 
radiation which will tend to smear out the fluctuations, or a process of
fluctuation generation which depends on radial position in the source.
Thus timing observations during absorption dips can give information
about the source region and may place constraints on its size. 
\end{abstract}



\section{Introduction}
Cygnus X-1 is well known as one of the best Galactic black hole
candidates, with a mass of the compact object in the range 4.8
- 14.7 $\rm {M_{\sun}}$ (Herrero et al. 1995). It is highly variable,
and on timescales of weeks and years shows at least two luminosity states: 
a Low State in which it spends most of its time, and a High State with much softer 
spectrum (see review by Tanaka and Lewin 1995).
There can also be transient reductions in X-ray flux lasting several
minutes: ie X-ray dips, during which the spectrum hardens, showing that
these are due to absorption (Kitamoto et al. 1984).
The X-ray spectrum of Cygnus X-1 is complex, consisting in the Low State
of a hard, underlying power law, a reflection component 
(Done et al. 1992), and a soft excess
(Ba\l uci\'nska \& Hasinger 1991). It was previously shown that in a {\it
Rosat} observation, $\rm {kT_{bb}}$ for the soft excess agreed well with 
the characteristic temperature of the inner accretion disc for the low 
luminosity state of the source, indicating that the soft excess was 
disc emission (Ba\l uci\'nska-Church et al. 1995). 
The source also shows rapid time variations on timescales from milliseconds to
10~s of seconds. This rapid aperiodic 
variability, often called flickering, was discovered in Cygnus X-1 by Oda et al. (1971). 
Since then there have been extensive studies of such variability in various 
black hole
candidates, and it is generally accepted that the flickering phenomenon
originates in the neighborhood of the inner accretion disc although the mechanism is not clear.
The strength of variability may be quantified in terms of the {\it fractional 
rms amplitude} $\rm {{\sigma} / I_x }$, and values of 30\% - 50\% are
typical of the Low State of Cygnus X-1 (Belloni \& Hasinger 1990). The evidence is that
this strong variability arises in the hard power law spectral component
that dominates the Low State of the source, whereas the soft spectral
component that dominates the High State shows little variability (van der
Klis 1995). In Cygnus X-1, previously no
major variation of the flickering has been detected during a particular
observation.
This {\it Letter} contains results of a 13 hr continuous observation of 
Cygnus X-1 by ASCA during which strong dipping took place, and we show 
that the amplitude of flickering decreased dramatically in dips.

\begin{figure}[h]
\leavevmode\epsffile{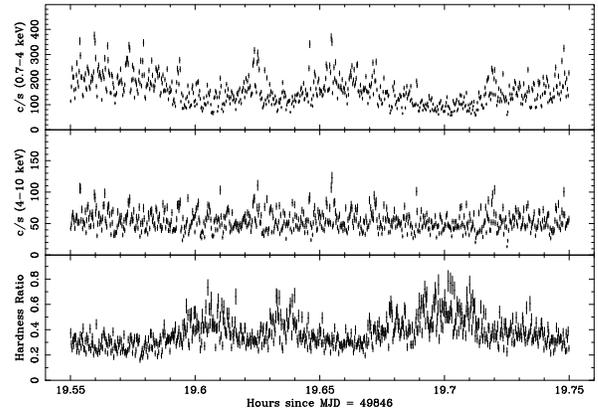}
\caption{GIS light curves in 1~s timebins
for the ASCA observation of Cyg X-1 of 1995 May 5th
in two energy band: 0.7 - 4.0 keV and 4.0 - 10.0 keV, together with
the hardness ratio formed by dividing the light curves. \label{fig1}}
\end{figure}

\section{Results}
\subsection {Spectral Analysis of Dip Evolution}  
The observation was made on 1995,
May 5th. by the satellite ASCA (Tanaka et al. 1994). Data were collected 
with the GIS detectors (Ohashi et al. 1996) and the SIS detectors, although 
only GIS data are discussed here.
At $\sim $ 19 $^{\rm h}$ UT dipping took place lasting approximately 40
minutes. 
The deadtime corrected light curve of the strongest dips are shown 
in Fig.1 in an expanded view with 1s 
binning.  The details of the dip light curves are given in two bands: 
0.7 -- 4.0 keV and 4.0 -- 10.0 keV, together with the hardness ratio (HR)
formed by dividing these.


Three dips can be seen lasting 2 to 3 min in each 
of which there is a strong increase in hardness ratio associated with the
photoelectric absorption taking place.  We have analysed the spectral
changes in dipping by dividing the data into 7 intensity bands including
non-dip emission, selected in a time interval including the strong 3rd
dip in Fig. 1, and non-dip data on each side of the dip. Spectral analysis
of the intensity-selected spectra in the band 0.7 -- 4.0 keV
were carried out using a blackbody to
represent the soft excess, and a power law to represent the hard
component, since in this band the reflection component
makes little contribution. Using the non-dip data, it was found that
the value of the blackbody temperature of the soft excess was
well constrained, with $\rm {kT_{bb}}$ = $\rm {0.13\pm 0.01}$ keV
(Ba\l uci\'nska-Church et al. 1996).
The dip data required a model in which partial
covering of the power law took place.
Dipping was seen to be due primarily to
absorption of the major power law component with $\rm {N_H}$
increasing from $\rm {8.9 ^{+18.5} _{-0.8}\cdot 10^{21}}$ H atom 
$\rm {cm^{-2}}$ in non-dip to 
$\rm {72 ^{+11} _{-9} \cdot 10^{21}}$ H atom $\rm {cm^{-2}}$ in the
deepest part of dipping, when the partial covering fraction was $\sim $
70\%. 
We were not able to constrain very well possible
changes in absorption of the blackbody taking place at energies below the 
ASCA band.

\begin{figure}[h]
\leavevmode\epsffile{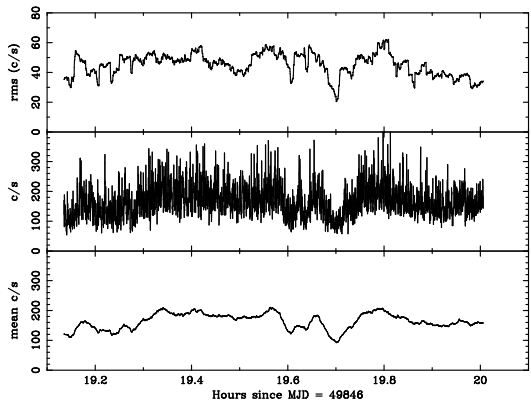}
\caption{(a) rms flickering amplitude evaluated in an 85~s timebin
running through the observation; (b) the unsmoothed X-ray light curve in
the band 0.7 - 4.0 keV; (c) running 85~s means of the light curve.
\label{fig2}}
\end{figure}

\smallskip 
\subsection {Phase of Dipping}  
The value of phase that we calculated for the observed
dipping using the ephemeris of Gies \& Bolton (1982) was 0.70 
$\rm {\pm 0.02}$.

Dipping has previously been seen at around phase
zero, consistent with inferior conjunction with the Companion. However we
cannot be definite about the actual phase because of the errors involved in
extrapolating this ephemeris to the date of the observation. This is
especially the case because of the possibility that the orbital period is
changing, as suggested by Ninkov et al. (1987) based on analysis of all data available
at that time. This will not be resolved until there is a new
determination of the ephemeris. It is important to decide this point,
since dipping at phase $\sim $0.7 would imply absorption in the stellar
wind of the Companion.

\begin{figure*}[t]
\leavevmode\epsffile{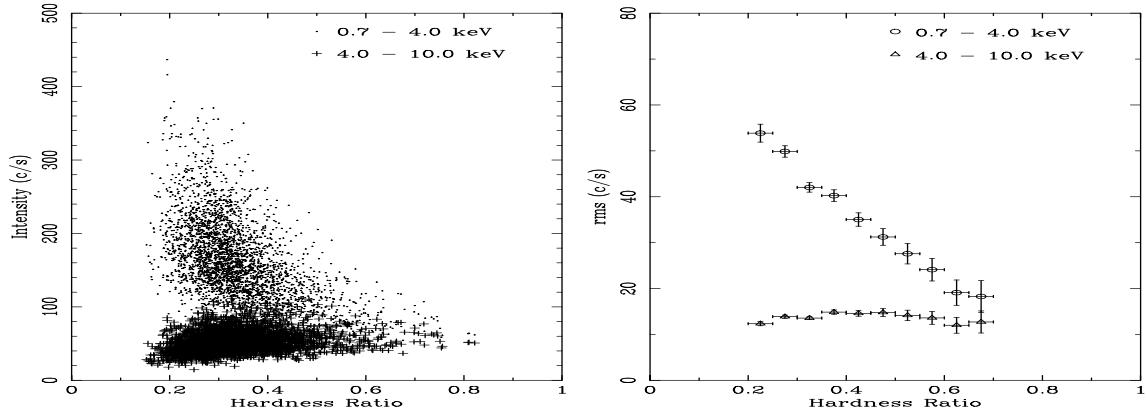}
\caption{(a) Intensity versus hardness ratio in the same two energy
bands as in Fig. 1, and (b) rms flickering amplitude in c/s versus
hardness ratio. 
\label{fig3}}
\end{figure*}

\subsection {Cessation of Flickering}  
The striking effect that flickering essentially 
stops in the deepest part of the dipping, is obvious in Fig. 1, particularly 
in the 3rd dip which is the strongest.  In the lower energy band 0.7 - 4.0 keV, 
it can be seen that flickering on timescales longer than the 1~s binning
has an amplitude of 200
c/s in non-dip emission, i.e. the soft X-ray intensity rises from 200 -- 400
c/s.  However, in dipping it can be seen that this amplitude decreases
considerably.  The effect is less obvious in the higher energy band 4.0 -
10.0 keV,  although there may be some small decrease in amplitude 
at the third dip.

\noindent
We have also plotted the data as a function of time in the observation
by evaluating the rms amplitude of the variability in 1~s bins in a 
running 85~s section of data, and also 
evaluating a running mean of the intensity in these 85~s sections.
This is shown in Fig. 2 for the lower energy band 0.7 - 4.0 keV. 
The middle panel shows the unintegrated light curve with 1~s binning,
and the lower panel shows the integrated average values in 85~s timebins.
There is some underestimation of the depth of dipping
in the averaged light curve, and also of the depth of dipping in the
rms amplitude plot as a result of smoothing over a long timebin.
However it is clear that there is a good correspondence between amplitude
and intensity.

\noindent
To demonstrate the effect more clearly, the data are replotted in Fig. 3a
and 3b.  In Fig. 3a the X-ray intensities in the soft band and the hard
band are plotted against hardness 
ratio, so that dipping corresponds to the high values of hardness ratio.
It is clear that the peak-to-peak variation in the X-ray intensity
falls sharply as dipping takes place. This is shown more clearly in
Fig. 3b in which the rms amplitude of intensity variation is 
plotted against hardness ratio for the two energy bands. This was produced
by evaluating the rms deviation of the variability of the data selected
within a narrow band of hardness ratio. Firstly, it can be seen that
the non-dip value of the amplitude of 60 c/s with a non-dip count rate of 
200 c/s in the low band gives a fractional rms amplitude of 30\% which
is quite typical for the Low State of Cygnus X-1.
There is a strong decrease of 
the amplitude at low energy, but with little change at high energies.
To show this more clearly, we plot rms amplitude against X-ray intensity
for the two energy bands in Fig. 4. Poisson noise is subtracted from the
rms, and also an approximate correction to the intensity is made for
the dust-scattered halo in the band 0.7 - 4.0 keV based on the work of
Predehl and Schmitt (1995). 
The scattered component is not expected to show
flickering as this will be smoothed out by the variable time delays. 
We estimate that the contribution to the intensity of the non-flickering
component in this energy band is $\sim $ 6\%.
The data in the low band are generally 
consistent with a simple relationship between the variability and
intensity which changes by almost a factor of 3 due to photoelectric
absorption.
However, the points at lowest dip intensity fall below a linear
relationship, and in terms of the $\rm {1\,\sigma }$ errors plotted, the
lowest 2 points are an average of $\rm {1.8\,\sigma}$ below the straight line. Thus the
presence of a departure from linearity is likely, but not proven.
In the high energy band
the amplitude remains constant and there is little change in intensity
since the increase in $\rm {N_H}$ in the dip has little effect in this
band. 


Thus, the
reduction in rms amplitude simply reflects the decreasing power law
intensity in the dips due to photoelectric absorption which is strong
in the low energy band, but has little effect in the high band.
Furthermore, if in the band 0.7 - 4.0 keV, the rms amplitude is divided
by the intensity, the fractional rms amplitude is approximately
constant, as expected from the approximate linearity of Fig. 4a, indicating
the simple relation between amplitude and intensity. 


\section{Discussion}
We have demonstrated for the first time the cessation of flickering 
during deep dipping in Cyg X-1.
The fractional rms amplitude of the variability is about 30\% in
the non-dip emission, a typical value for the Low State.
In the energy band 0.7 -- 4.0 keV, there are two spectral components:
the soft excess blackbody and the power law. Taking typical parameters
for the soft excess, as determined from ASCA and {\it Rosat}
(Ba\l uci\'nska-Church et al. 1995), we can estimate that this
contributes only 8\% of the count rate in the band 0.7 - 4.0 keV. 
Thus it is clear that the power law {\it must} be involved in the
variability, and the linearity implies that the variability originates in 
the power law emission region, not external to this.

\begin{figure*}[t]
\leavevmode\epsffile{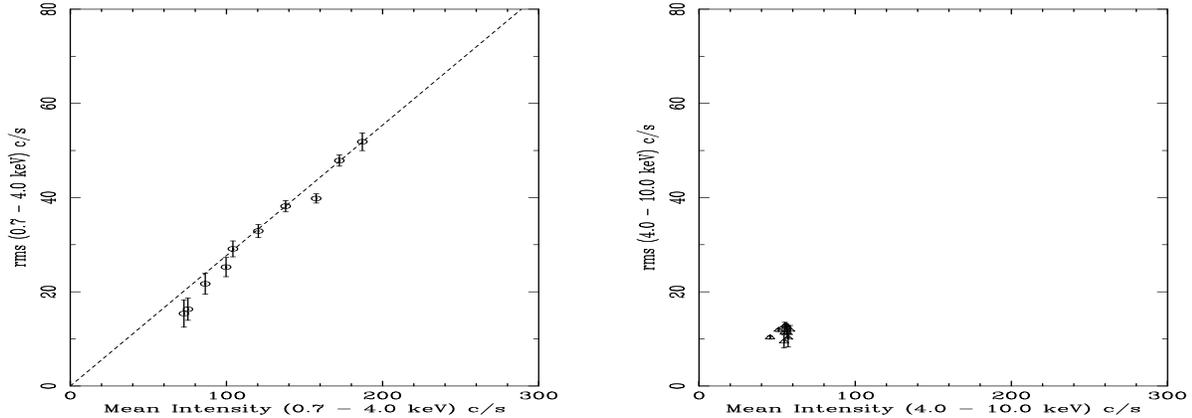}
\caption{Rms amplitude versus x-ray intensity (a) in the band 0.7 -
4.0 keV and (b) in the band 4.0 - 10.0 keV. 
\label{fig4}}
\end{figure*}

\smallskip
Secondly, we have shown that the change in the strength of variability
is simply related to the total intensity (dominated by the power law
component) which changes by almost a factor of three
in the energy band 0.7 - 4.0 keV due to photoelectric absorption.
This is consistent with the simple expectation that, if the intensity falls 
by a given factor due to photoelectric absorption,
then absorption will reduce the variability by the same factor.
The departure from linearity in Fig. 4, if substantiated, may reveal futher
information about the source, and so we discuss below possible reasons
for this non-linearity. Firstly, we have not
attempted to correct the data for the contribution to the flickering 
or to the intensity of the soft excess. The flickering in this component 
may well be very much less than in the power law. However, because the
soft excess, with $\rm {kT_{bb}}$ = 0.13 keV, contributes only at the
lowest energies in the ASCA band (below 1.5 keV),
it will be totally removed by absorption at an early stage of dipping.
Thus, at total intensities below 150 c/s in Fig. 4, only
the power law component remains, and the plot becomes a plot of power
law rms {\it versus} power law intensity, and so the curvature below
100 c/s is unlikely to be related to the soft excess spectral
component. This depends only on the reasonable assumption that this
component originates in the central part of the source and so is covered
by the absorber during dipping.
There are however, effects which can lead to curvature in the plot
at low values of intensity. Firstly, the geometry of the absorber
causing the absorption dip may be such as to cover the central part of
the emission regions only, as suggested by our result that the partial
covering fraction rises to only 70\% in the deepest parts of dips.
If the process generating the flickering were to fall off with distance from
the center of the emission region, then the uncovered part of the emission
will have reduced flickering in dipping, leading to curvature in the plot. 


Secondly, electron scattering close to the source
region could produce an effect. If part of the power law X-ray emission
gets scattered, then the fast variability will tend to get smeared out.
In 1~s timebins, we are determining the rms of longer-lasting shots
more than $\sim $ 200 ms in length, although the exact timescale depends
on the shot profile. 
Kitamoto et al. (1984) argued from the timescales of ingress to and
egress from dipping in high time resolution TENMA observations, 
that the major source region was smaller than $\rm {4\cdot
10^8}$ cm. Scattering in this region would introduce a
variable delay of up to 13 ms which would cause some reduction of flickering
in the scattered component for the longer shots. However a region of
size $\rm {4\cdot 10^8}$ cm is very small in comparison with typical
sizes of accretion disk coronae, for example, and if the scattering region
was only as large as $\rm {4\cdot 10^9}$ cm, there would be a delay of
130 ms which would cause a major reduction of flickering in the 
scattered component. Then, as the source is only 70\% covered in deep
dipping, this component would not be completely absorbed but would have 
reduced variability, leading to curvature in the plot of rms {\it versus} 
intensity. 

\smallskip
\par Analysis of the present data constitutes only a first attempt at investigating
the results of absorption on the fast aperiodic variability. In principle,
this can provide information on the origins of the variability,
on the relation between the spectral
components and the variability, on the extent to which the source is
covered by the absorber, and on possible effects such as electron
scattering. Further more  detailed work may resolve some of these aspects.



\begin{references}
\reference{}
Ba\l uci\'nska M. and Hasinger G., 1991, A\&A {\bf 241}, 439. 
\reference{}
Ba\l uci\'nska-Church M., Belloni T., Church M. J. and Hasinger G., 1995,
A\&A {\bf 302}, L5.
\reference{}
Ba\l uci\'nska-Church M., Church M. J., Ueda Y. et al., 1996,
Proc. of Conf. on X-ray Imaging and Spectroscopy of Cosmic Hot Plasmas,
March 1996, Waseda University, Tokyo.
\reference{}
Belloni T. and Hasinger G., 1990, A\&A {\bf 227}, L33.
\reference{}
Done C., Mulchaey J. S., Mushotzky R. F. and Arnaud K. A., 1992, ApJ
{\bf 395}, 275.
\reference{}
Gies D. R. and Bolton C. T., 1992, ApJ {\bf 260}, 240.
\reference{}
Herrero A., Kudritzki R. P., Gabler R., Vilchez J. M. and Gabler A., 1995,
A\&A {\bf 297}, 556.
\reference{}
Kitamoto S., Myamoto S., Tanaka Y. et al., 1984, PASJ {\bf 36}, 731.
\reference{}
Ninkov Z., Walker G. A. H. and Yang S., 1987, ApJ {\bf 321}, 425.
\reference{}
Oda M., Gorenstein P., Gursky H. et al.,
1971, ApJ {\bf 166}, L1.
\reference{}
Ohashi T., Ebisawa K., Fukazawa Y. et al.,
1996, PASJ {\bf 48}, 157.
\reference{}
Predehl P. and Schmitt J. H. M.,  1995, A\&A {\bf 293}, 889.
\reference{}
Tanaka Y., Inoue H. and Holt S. S., 1994, PASJ {\bf 46}, L37.
\reference{}
Tanaka Y. and Lewin W. H. G., 1995, in ``X-ray Binaries'', Cambridge
Astrophysical Series vol. 26, Cambridge University Press.
\reference{}
Van der Klis M., 1995, in `` X-Ray Binaries'', Cambridge Astrophysical
Series vol. 26, Cambridge University Press.
\end{references}
\end{document}